\documentclass[doublecol]{epl2}
\usepackage{amsmath}
\usepackage{graphicx}
\usepackage{amssymb}
\usepackage{color,soul}
\usepackage[T1]{fontenc}
\usepackage{bm}
\usepackage{ae,aecompl}
\usepackage{nameref}
\expandafter\let\csname equation*\endcsname\relax
\expandafter\let\csname endequation*\endcsname\relax
\title{Dynamical attractors of memristors and their networks}
\author{Y. V. Pershin\inst{1} \and V. A. Slipko\inst{2}}
\shortauthor{Y. V. Pershin and V. A. Slipko}
\institute{
  \inst{1} Department of Physics and Astronomy, University of South Carolina, Columbia, South Carolina 29208, USA\\
  \inst{2} Institute of Physics, Opole University, Opole 45-052, Poland
}

\abstract{
It is shown that the time-averaged dynamics of memristors and their networks periodically driven by alternating-polarity pulses  may converge to fixed-point attractors. Starting with a general memristive system model, we derive basic equations describing the fixed-point attractors and investigate  attractors in the dynamics of ideal, threshold-type and second-order memristors, and memristive networks. A memristor \textit{potential function} is introduced, and it is shown that in some cases the attractor identification problem can be mapped to the problem of potential function minimization. Importantly, the fixed-point attractors may only exist if the function describing the internal state dynamics depends on an internal state variable. Our findings may be used to tune the states of analog memristors, and also be relevant to memristive synapses subjected to forward- and back-propagating spikes.}

\begin{document}

\maketitle

\section{Introduction} \label{sec:Intro}

Setting the state of memristive systems~\cite{chua76a} (memristors) is an important step in a number of their applications including neural networks~\cite{alibart2013pattern} and various analog circuits~\cite{pershin09d}. Generally, this task can be performed with and without feedback. Feedback-based schemes~\cite{Alibart12a,Berdan12a} are more precise, as they involve the use of adaptive pulse sequences and device state monitoring. However, extra circuitry necessary to implement such schemes adds additional unwanted complexity to the design and requires extra space. To set the memristor state without feedback, the memristor can be placed first into its ``on'' or ``off'' state (low- or high-resistance state, respectively) and then set into the desired state by application of voltage/current pulses. However, a better accuracy is achieved employing the voltage divider effect~\cite{Kim15a,kim2016voltage,vourkas2017exploring}, which eliminates the errors associated, e.g., with the variability of the "on" and "off" states of memristors.

The present paper introduces a novel non-feedback approach to the memristor state initialization. One of our main findings is the existence of fixed-point attractors of driven memristors and their networks. We show that under appropriate conditions, memristor can be placed into the desired (attractor) state from any initial state by an appropriate sequence of pulses. Another important result is that some of driven memristors can be described in terms of a potential function whose minima correspond to equilibrium points.  Irrespective of the existence of potential function, the fixed-point attractors can be used for high-precision tuning of analog state of memristors and memristive networks. We also anticipate that the memristor attractors may be relevant to some neural networks in which, for instance, memristive synapses are subjected to forward- and back-propagating spikes. Our findings are different from the previously discussed attractors in Chua's~\cite{Chen2015,kengne2015periodicity} and other~\cite{bao2016coexisting} circuits as these circuits contain non-resistive components. Moreover, we emphasize that the attractors considered in this work are possible only with certain types of memristors/networks.

\begin{figure*}[t]
\centering (a)\includegraphics[width=.20\columnwidth]{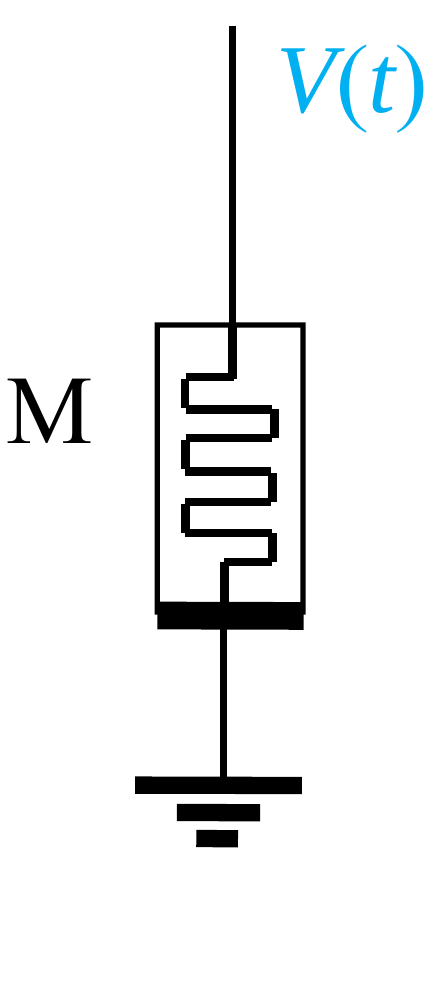}\;\;\;(b)\includegraphics[width=.80\columnwidth]{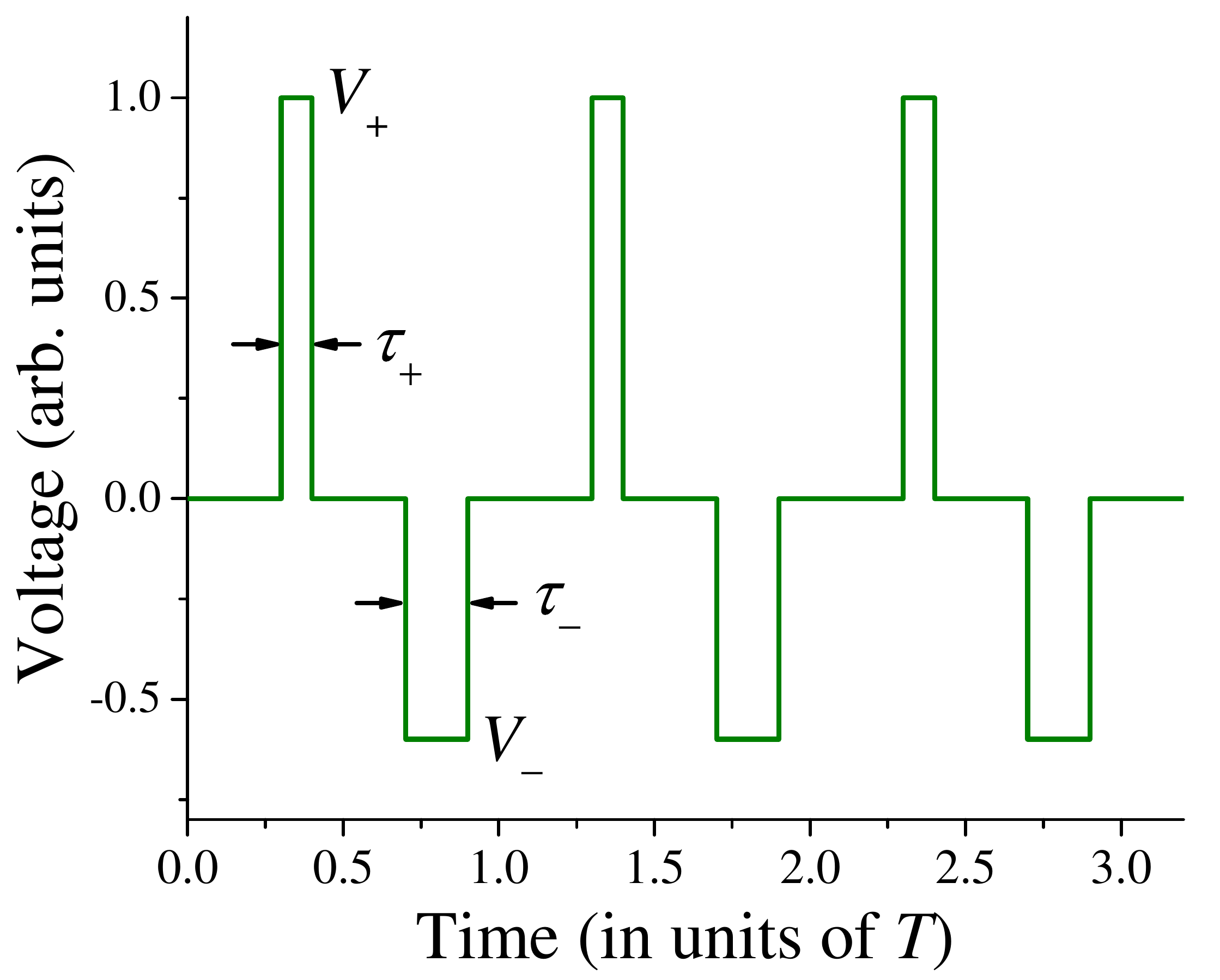}\;\;\;
(c)\includegraphics[width=.80\columnwidth]{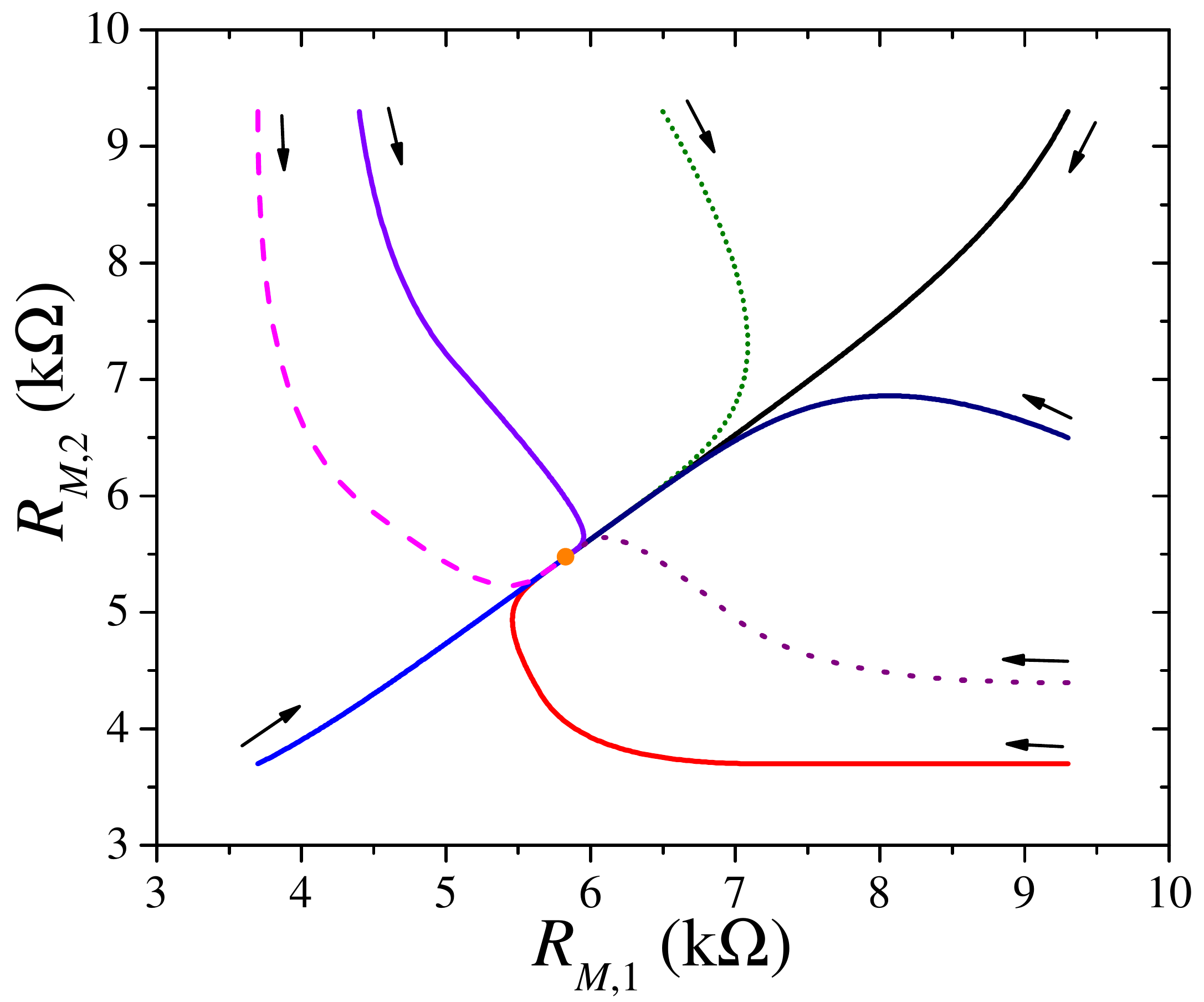}
\caption{ (a) Schematics of the circuit  considered in this work. Here, M denotes either a single memristor or two-terminal network of memristive/resistive devices. (b) Pulse sequence $V(t)$ applied to the memristor. In (b), $V_+>0$ and $V_-<0$ are the amplitudes of positive and negative pulses, $\tau_+$ and $\tau_-$ are their widths, and $T$ is the period. In this paper, $V_+>0$, $V_-<0$,
$\tau_+$, and $\tau_-$ are external control parameters. (c) Attractor of
a resistor-two memristors network. The attractor point is indicated by the dot. The arrows depict the direction of evolution. For additional details, see the subsection {\it Resistor-two memristors network}.}
\label{fig:1}\end{figure*}

Figure \ref{fig:1}(a) and (b) present the schematics of circuit and pulse sequence we are dealing with in this study. Specifically, we consider a voltage-controlled memristive system directly connected to a voltage source (Fig. \ref{fig:1}(a)). It is assumed that the applied pulse sequence (Fig. \ref{fig:1}(b)) consists of narrow square-shaped pulses (spikes) such that the change of the internal state variable $\bm{x}$ of memristor with each pulse is small ($\bm{x}$ is defined below). In the simplest case there are two opposite-polarities pulses per period, however, more general cases are also considered. An example of attractor point is demonstrated in Fig. \ref{fig:1}(c). This plot shows that the states of memristors in a certain resistor-two memristors network (details are given below) evolve towards the same steady state (attractor point), for a wide variety of initial conditions of the network.

For the discussion below we will need the definition of voltage-controlled memristive systems. They are a class of two-terminal devices with memory defined by~\cite{chua76a}
\begin{eqnarray}
I&=&R^{-1}_M\left( \bm{x}, V_M \right) V_M, \label{eq:1} \\
\dot{\bm{x}}&=&\bm{f}\left(\bm{x}, V_M \right), \label{eq:2}
\end{eqnarray}
where $I$ and $V_M$ are the current through and voltage across the system, respectively, $R_M\left( \bm{x}, V_M\right)$ is the memristance (memory resistance), $\bm{x}$ is an $n$-component vector of internal state variables and $\bm{f}\left(\bm{x}, V_M\right)$ is a vector-function. Here, the bold font is used to denote
vectors and normal font is used to denote scalar quantities. It should be mentioned that the main formulae derived below can be easily adapted to describe current-controlled memristive systems~\cite{chua76a} driven by current pulses.


This paper is organized as follows. To simplify the presentation, we start with memristors based on a single internal state variable $x$. We derive the conditions necessary for an attractor and introduce the memristor potential function. In Sec. {\it Multivariable Memristors}, these concepts are generalized to the case of multivariable memristors (their internal state is a vector $\bm{x}$). Examples of fixed-point attractors are considered in Sec. {\it Examples}. The manuscript is concluded
with some concluding remarks.

\section{Memristors described by a single internal state variable} \label{sec:2}

Consider a memristor~\footnote{Note that any two-terminal network of memristive/resistive components can be considered as an effective memristor~\cite{chua71a}.} subjected to a periodic train of narrow alternating-polarity pulses shown in Fig. \ref{fig:1} (b). Our interest is to understand whether such pulse train can place the memristor into a certain state (different from its ``on'' and ``off'' states) from any initial state. In other words, we are asking if there exists a fixed-point attractor (or attractors) in the time-averaged dynamics of driven memristors.

For pedagogical reasons we first consider memristors described by a single internal state variable $x$. Its dynamics is given by Eq. (\ref{eq:2}), which makes the basis for our study. Under the action of positive ($+$) and negative ($-$) pulses (see Fig. \ref{fig:1}(b)), the  changes of the internal state variable $x$ are given by $\Delta x_+=f(x,V_+)\tau_+$ and $\Delta x_-=f(x,V_-)\tau_-$, respectively. If a fixed-point attractor exists in the time-averaged dynamics over the pulse period $T$ then, at the attractor point $x_a$, $\Delta x_++\Delta x_-=0$ or
\begin{equation}
f(x_a,V_+)\tau_++f(x_a,V_-)\tau_-=0 .  \label{eq:3}
\end{equation}
Next, consider a state displaced by a small $\delta x$ from the equilibrium point assuming at the same time that
 $|\Delta x_\pm|\ll|\delta x| $. Then, in the displaced state,
\begin{equation}
\Delta x_++\Delta x_-=f(x_a+\delta x,V_+)\tau_++f(x_a+\delta x,V_-)\tau_-.  \label{eq:4}
\end{equation}
Expanding Eq. (\ref{eq:4}) with respect to small $\delta x$ one finds that the internal state variable $x$ will drift towards the attractor if
\begin{equation}
\left.\frac{\partial f(x,V_+)}{\partial x}\right|_{x=x_a}\tau_++ \left.\frac{\partial f(x,V_-)}{\partial x}\right|_{x=x_a}\tau_-<0.  \label{eq:5}
\end{equation}
To summarize, the point $x=x_a$ is a fixed-point attractor, if, at this value of $x$, Eq. (\ref{eq:3}) and inequality (\ref{eq:5}) are simultaneously satisfied.

Moreover, one can notice that the left-hand side of Eq. (\ref{eq:5}) is the derivative of the left-hand side of Eq. (\ref{eq:3}) with respect to $x$. Considering $[f(x,V_+)\tau_++f(x,V_-)\tau_-]$ as a force and $x$ as a coordinate, one can introduce a memristor \textit{potential}
\textit{function}
\begin{equation}\label{eq:6}
U(x)=-\int \left[ f(x,V_+)\tau_++f(x,V_-)\tau_- \right] \textnormal{d}x.
\end{equation}
Analyzing Eq. (\ref{eq:6}), one can notice that the minima of $U(x)$
correspond to fixed-point attractors in driven memristors. Indeed, Eq. (\ref{eq:3}) coincides with $\textnormal{d} U(x)/\textnormal{d} x=0$ and Eq. (\ref{eq:5}) matches $\textnormal{d}^2 U(x)/\textnormal{d} x^2>0$.

We emphasize that Eqs. (\ref{eq:3}), (\ref{eq:5}), and (\ref{eq:6}) are also valid for current-controlled memristors driven by current pulses with the appropriate replacement of voltage by current. An additional point to emphasize is that when the state of memristor is in the basin of attraction, the total dynamics is given by a superposition of  a slow drift towards the fixed point (on the time scale larger than the pulse period) and fast oscillations (on the time scale of  the period).

\section{Multivariable Memristors}  \label{sec:3}

Now we generalize the previous section to the case of memristors described by $n$ internal state variables and $m$ pulses per period. Assuming sufficiently short pulses, the change of the vector $\bm{x}$ within the period $T$ can be written as
\begin{equation}\label{eq:7}
  \Delta\bm{x}=\sum\limits_{k=1}^{m} \Delta\bm{x}_k=\sum\limits_{k=1}^{m} \bm{f}(\bm{x},V_k)\tau_k.
\end{equation}
Here, the external control parameters $V_k$ and $\tau_k$ are defined similarly to $V_\pm$ and $\tau_\pm$ in Fig.~\ref{fig:1}.
Analogously to Eq. (\ref{eq:3}) above, at the attractor point $\bm{x}_a$ we require that $\Delta\bm{x}=0$ or
\begin{equation}\label{eq:8}
  \sum\limits_{k=1}^{m} \bm{f}(\bm{x}_a,V_k)\tau_k=0.
\end{equation}
Clearly, if Eq. (\ref{eq:8}) is satisfied then the memristor will remain in the same internal state after the time interval~$T$.

Let us introduce a vector function
\begin{equation}\label{eq:9}
 \bm{\Phi} (\bm{x})= \sum\limits_{k=1}^{m} \bm{f}(\bm{x},V_k)\tau_k.
\end{equation}
Then, Eq. (\ref{eq:8}) can be represented as
\begin{equation}\label{eq:10}
  \Delta\bm{x}=\bm{\Phi} (\bm{x}_a)=0.
\end{equation}
Consider now a small (but finite) displacement of the internal state from the attractor point $\bm{x}_a$ to $\bm{x}_a+\delta \bm{x}$. After the time interval $T$, the internal state will be changed by
\begin{equation}
\label{eq:11}
  \Delta\bm{x}=\bm{\Phi} (\bm{x}_a+\delta\bm{x})=
 \left.(\delta \bm{x}\cdot\nabla)\bm{\Phi} (\bm{x})\right|_{\bm{x}=\bm{x}_a}. \;\;\;
\end{equation}

The point $\bm{x}_a$ is an attractor  if after the time interval $T$ the final memristor state is closer to $\bm{x}_a$ compared to the initial  state $\bm{x}_a+\delta \bm{x}$. Mathematically, this condition can be written as
\begin{equation}\label{eq:12}
  \| \delta \bm{x}+ \Delta \bm{x} \| <  \| \delta \bm{x} \|
\end{equation}
for any small $\delta \bm{x}\neq 0$. Here, $\| ... \|$ denotes the distance in the configuration space of memristor. The Euclidean norm is the first choice for $\| ... \|$, although, strictly speaking, the physics of the internal state should be taken into account when choosing the norm. Below, we will use
\begin{equation}\label{eq:13}
  \| \bm{x} \|^2 = x_1^2+x_2^2+...+x_n^2.
  \end{equation}

Let us first analyze the condition  (\ref{eq:12}) for the existence of attractor at $\bm{x}_a$ in its most general form, which, taking into account Eq. (\ref{eq:11}), can be presented as
\begin{equation}\label{eq:14}
  \| (1+\hat{F}(\bm{x}_a))\delta \bm{x} \| <  \| \delta \bm{x} \|,
\end{equation}
where $\hat{F}(\bm{x})$ is a linear operator (represented by $n\times n$ matrix) acting on $n$ components of $\delta \bm{x}$ according to
\begin{equation}\label{eq:15}
  \hat{F}(\bm{x})\delta \bm{x} =
  (\delta \bm{x}\cdot\nabla)\bm{\Phi} (\bm{x}),
\end{equation}
i.e. $F_{ij}=\partial \Phi_i(\bm{x})/ \partial x_j$.

It follows from inequality (\ref{eq:14}) that the norm of the linear operator $1+\hat{F}$ should be less than $1$, namely,
\begin{equation}\label{eq:16}
  \| 1+\hat{F}(\bm{x}) \| < 1.
\end{equation}
Therefore, the linear operator $1+\hat{F}(\bm{x})$ must be a contraction operator at the attractor point $\bm{x}_a$. There are several known properties of such operators. In particular, all eigenvalues $\lambda$ of $\hat{F}(\bm{x})$ should lay inside the unit circle centered at $-1$:  $|1+\lambda|<1$.

It should be noted that Eq. (\ref{eq:10}) alongside with inequality (\ref{eq:16}) completely determine the attractor points $\bm{x}_a$ in the most general form. For sufficiently short pulses, when $ \hat{F}(\bm{x})\propto \tau$ (see Eq. (\ref{eq:9})), we can neglect the quadratic in $\tau$ terms  in  (\ref{eq:14}). In this very important case, the inequality (\ref{eq:14}) reduces to
\begin{equation}\label{eq:17}
\sum\limits_{i,j=1}^{n} F_{ij}\delta x_i \delta x_j<0.
\end{equation}
This simply means that the symmetrised matrix
\begin{equation}\label{eq:18}
 \tilde{F}_{ij}=-\frac{(F_{ij}+F_{ji})}{2}=-\frac{1}{2}\left(\frac{\partial \Phi_i(\bm{x})}{\partial x_j}+\frac{\partial \Phi_j(\bm{x})}{\partial x_i}\right)
\end{equation}
should be positive-definite at the attractor points $\bm{x}_a$, what can be checked using, for instance, the
Sylvester's criterion.

We note that the potential function can be introduced only for those multivariable  memristors with $\bm{f}(\bm{x},V)$ satisfying
\begin{equation}\label{eq:22}
\frac{\partial f_i(\bm{x},V)}{\partial x_j}=\frac{\partial f_j(\bm{x},V)}{\partial x_i}.
\end{equation}
This class of memristors can be called as \textit{conservative memristive systems}.

Assuming that Eq. (\ref{eq:22}) is satisfied, it is convenient to split the potential function into components $u(\bm{x},V_k)$ as
\begin{equation}\label{eq:21}
 U(\bm{x})= \sum\limits_{k=1}^{m} u(\bm{x},V_k)\tau_k,
\end{equation}
where $u(\bm{x},V_k)$ are defined by
\begin{equation}\label{eq:19}
f_i(\bm{x},V_k)=-\frac{\partial u(\bm{x},V_k)}{\partial x_i}.
\end{equation}
Now it becomes clear that Eq. (\ref{eq:22}) represents the fact that the mixed partial derivatives of $u(\bm{x},V_k)$ are equal.

It follows from Eqs. (\ref{eq:9}) and (\ref{eq:18}) that for conservative memristive systems
\begin{equation}\label{eq:20}
 \tilde{F}_{ij}=-F_{ij}=\frac{\partial^2}{\partial x_i \partial x_j} \sum\limits_{k=1}^{m} u(\bm{x},V_k)\tau_k.
\end{equation}
One can notice that in the potential case the conditions for the attractor existence given by Eqs. (\ref{eq:10}),  and (\ref{eq:17}), with Eqs.   (\ref{eq:19}) and (\ref{eq:20}) taken into account, coincide with the conditions for the potential function minimum.


\section{Examples} \label{sec:4} \subsection{Ideal memristors}
The ideal memristors~\cite{chua71a} are hypothetical devices~\cite{vongehr2015missing,pershin18a} described by
\begin{equation}\label{eq:80}
  V_M(t)=R_M(q)I(t),
\end{equation}
where $q$ is the charge that has flowed through them from an initial moment of time. According to Eq. (\ref{eq:80}) definition, the internal state variable of ideal memristors is the charge and thus the corresponding Eq. (\ref{eq:2}) can be formulated as $\dot x=I$, so that $f=f(I)$.

There are now two possibilities to consider. In the case of ideal memristors subjected to current pulses, the current-controlled form of Eq. (\ref{eq:2}) should be used. Clearly, there are no attractor points in the dynamics of such memristors as their potential function (as follows from Eq. \ref{eq:6})) is linear in $x$. In the case of ideal memristors subjected to voltage pulses (so that the voltage is used as the external control parameter), Eq. (\ref{eq:2}) of ideal memristors can be written as
\begin{equation}\label{eq:81}
  \dot x=I=\frac{V_M}{R(x)}.
\end{equation}
Eq. (\ref{eq:81}) represents a separable model, where the function $f(x,V_M)=g(x)\cdot h(V_M)$. Consider now Eq. (\ref{eq:3}). Clearly, there are  no attractor solutions of Eq. (\ref{eq:3}) when $f(x,V_M)$ is separable (this statement can be straightforwardly verified by direct substitution).

\subsection{Threshold-type memristor}

Next we consider threshold-type memristors described by~\cite{pershin09b}
\begin{equation}\label{eq:30}
  \dot{x}=\left\{
                \begin{array}{ll}
                  k\left( V_M-V_{on} \right) ,  \;\;\;\;\;\;\;\;\; \;\;\;\;\;\; 0<V_{on}<V_M\\
                  0, \;\;\;\;\;\;\;\;\;\;\;\;\;\;\;\;\;\;\;\;\;\;\;\;\;\;\;\; \;\; \;\; V_{off}<V_M<V_{on} \;\;\;\;\;\; . \\
                  k \left(V_M - V_{off} \right),  \;\;\;\;\;\;\;\;\;\;\;\;\;\;   V_M<V_{off}<0
                \end{array}
              \right.
\end{equation}
Here, $k$ is the constant and $V_{on}$ and $V_{off}$ are the positive and negative threshold voltages, respectively. It is also assumed that $x$ is bound to the region between 0 and 1, and the memristance is a linear function of $x$,
\begin{equation}\label{eq:33}
  R_M(x)=R_{off}+x\left( R_{on}-R_{off}\right),
\end{equation}
where $R_{off}$ and $R_{on}$ are the high- and low-resistance states of memristor (the ``off'' and ``on'' memristor states, respectively).
This model presents a significant interest since  the threshold-type behavior is quite common in experimental memristive devices~\cite{pershin11a}.

\begin{figure}[tb]
\centering (a) \includegraphics[width=.9\columnwidth]{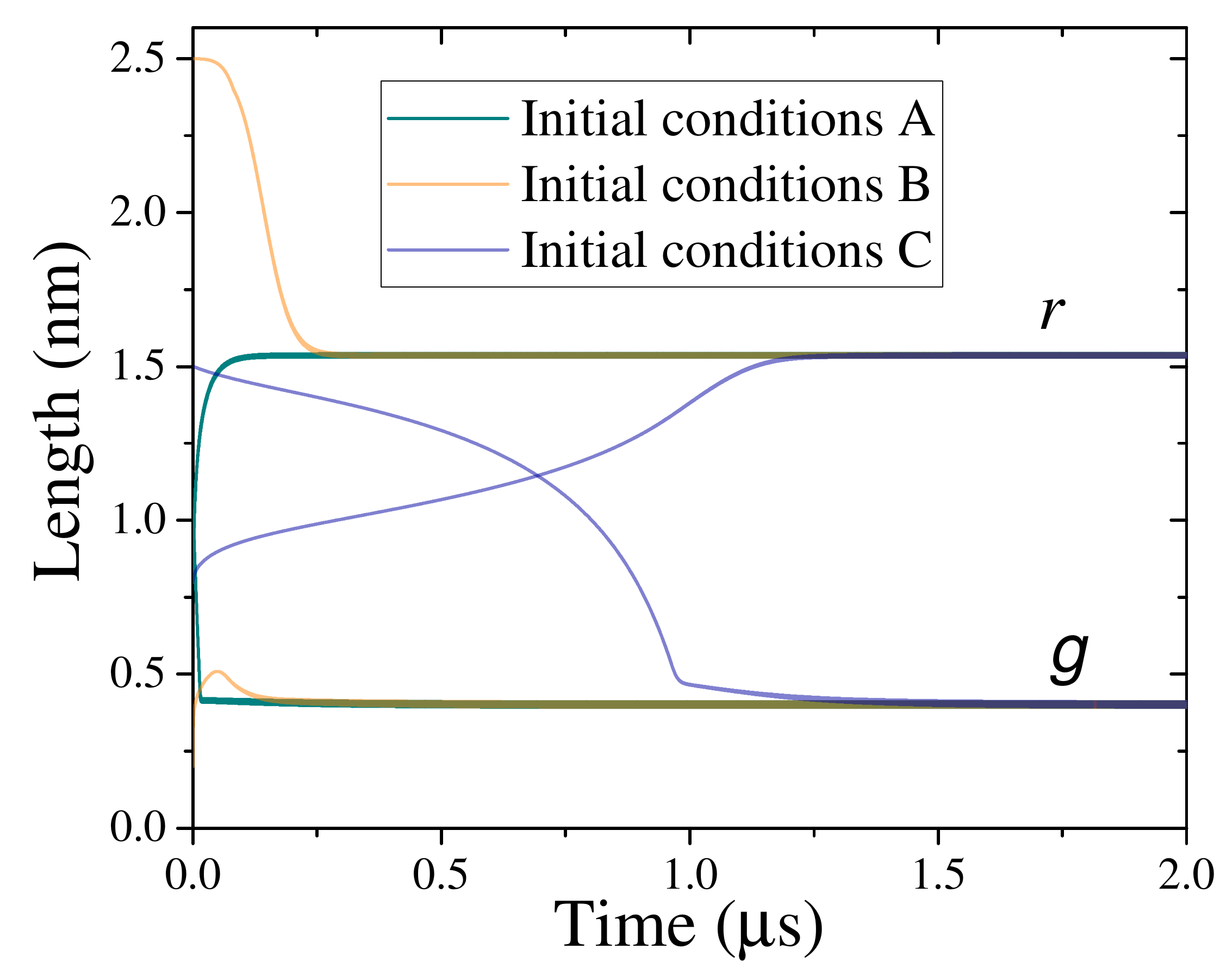} \\
(b) \includegraphics[width=.9\columnwidth]{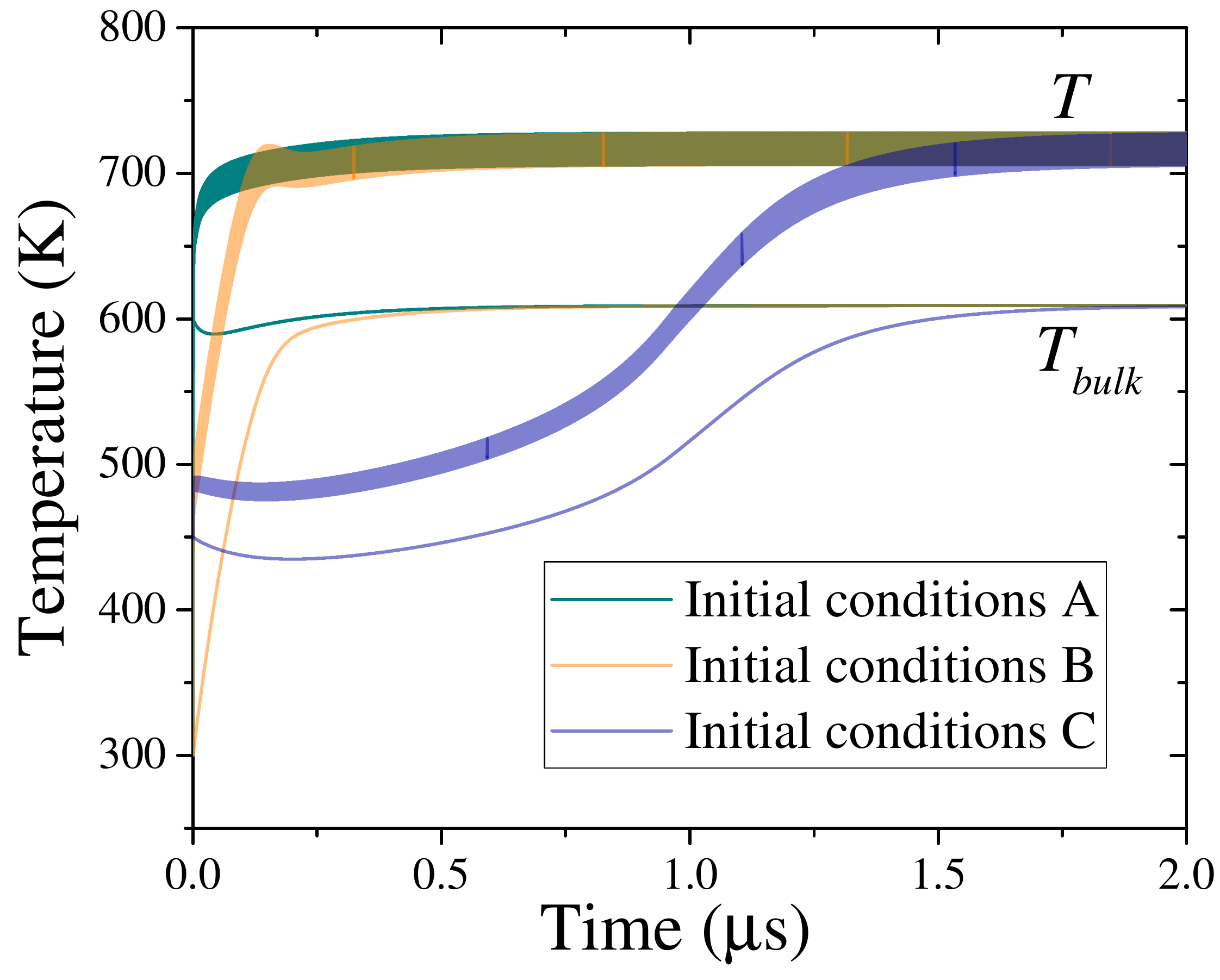}
\caption{Dynamics of internal state variables in a second-order memristor for several initial conditions.
(a) Temporal evolutions of the filament radius $r$ and depletion gap length $g$. (b) Temporal evolutions of the filament temperature $T$ and bulk temperature $T_{bulk}$.  The following initial conditions (chosen arbitrary) were used: (A) $r=0.8$~nm, $g=1$~nm, $T=300$~K, $T_{bulk}=600$~K; (B) $r=2.5$~nm, $g=0.2$~nm, $T=300$~K, $T_{bulk}=300$~K; (C) $r=0.8$~nm, $g=1.5$~nm, $T=450$~K, $T_{bulk}=450$~K.}
\label{fig:6}
\end{figure}

Similarly to the case of ideal memristors subjected to current pulses, the function describing the internal state dynamics of threshold-type memristor does not depend on $x$ (see the right-hand side of Eq. (\ref{eq:30})). Therefore, the potential function of threshold-type  memristors is a  linear function of $x$, meaning the absence of attractors.

\subsection{Second-order memristor}
It is important to evaluate the possibility of attractor dynamics in real experimental devices.
For this purpose, we will consider so-called second-order memristors, that were implemented recently using
certain tantalum oxide-based structures~\cite{kim15b}. It was shown that their dynamics can be
described by two sets of internal state variables~\cite{kim15b}. The first set
includes geometric parameters of a conducting filament~\cite{ielmini2011modeling} -- its radius, $r$, and depleted gap length, $g$. The second set of variables consists of two temperatures: the filament temperature, $T$, and the effective temperature in
the bulk of the device outside of the filament, $T_{bulk}$. As a compact model of such memristor involves four internal state variables,
according to the standard terminology~\cite{chua76a}, such a memristor should be classified as a fourth-order voltage-controlled memristive system. In what follows, however, we will use the original terminology ("the second-order memristor") used by the cited authors~\cite{kim15b}.

\begin{figure}[tb]
\centering \includegraphics[width=.9\columnwidth]{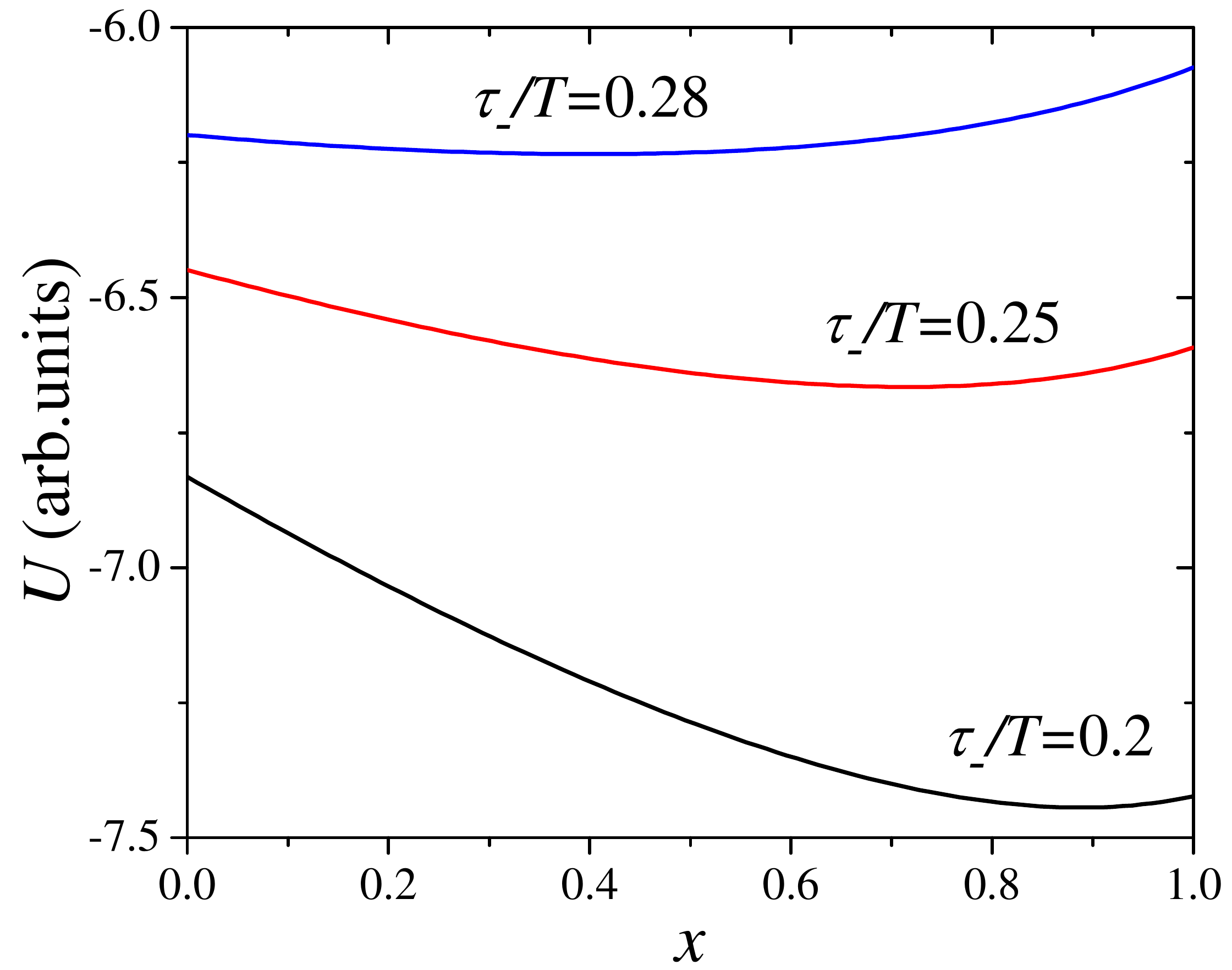}
\caption{Potential function of the resistor-memristor network (Eq. (\ref{eq:35})). The position of minimum corresponds to the attractor point. This plot was obtained using the following parameter values: $R=2$~k$\Omega$, $R_{on}=2$~k$\Omega$, $R_{off}=10$~k$\Omega$, $V_{on}=1$~V, $V_{off}=-0.7$~V,
$V_{+}=-V_-=2.2$~V, $\tau_+/T=0.4$. The curves were shifted for clarity.}
\label{fig:2}
\end{figure}

The dynamical attractors of second-order memristors were investigated using a SPICE model formulated in the Supplementary Information of Ref.~\cite{kim15b}. The only modification to the original SPICE model was the removal of a conditional statement in the definition of B1 current source. The transient dynamics of
a single second-order memristor was simulated in LTspice XVII. In our simulations, the second-order  memristor was subjected to a square voltage signal with $V_+=2$~V, $V_-=-1$~V, $\tau_+=0.2$~ns, and
 $\tau_-=0.1$~ns. The memristor dynamics from initial conditions A, B, and C (specified in the caption of Fig. \ref{fig:6}) was simulated for $2$ $\mu$s.

Figure \ref{fig:6} presents results of our simulations. According to Fig.~\ref{fig:6}, the second-order memristor evolves to the same final state for each set of initial conditions used in our simulations. Such a behavior is a good indicator of the attractor point.
Additional insights can be gained from an analytical analysis of the second-order memristor model, which is, however, beyond the scope of
this paper.

\subsection{Resistor-memristor network}


\begin{figure}[tb]
\centering (a) \includegraphics[width=.9\columnwidth]{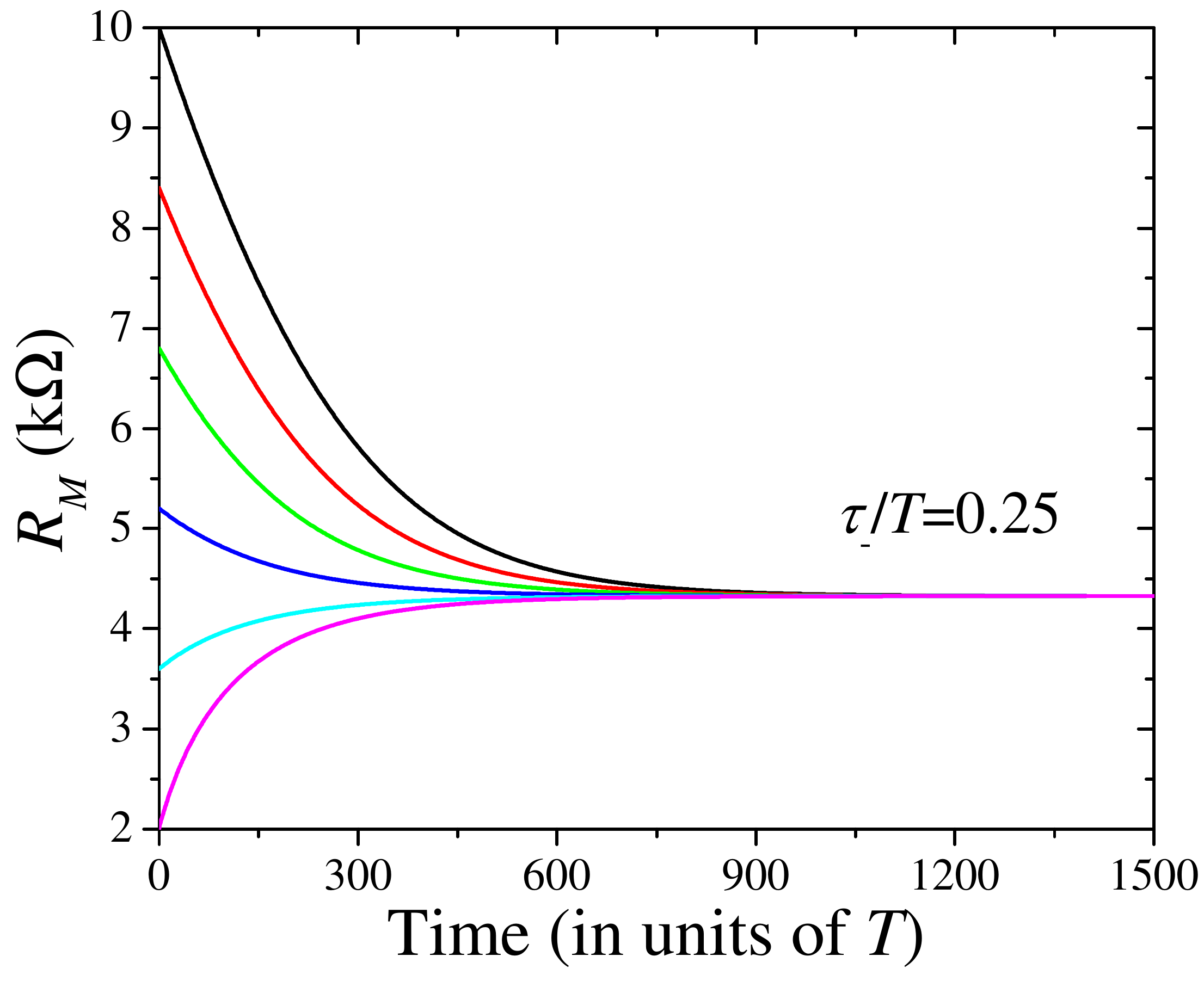} \\ (b) \includegraphics[width=.9\columnwidth]{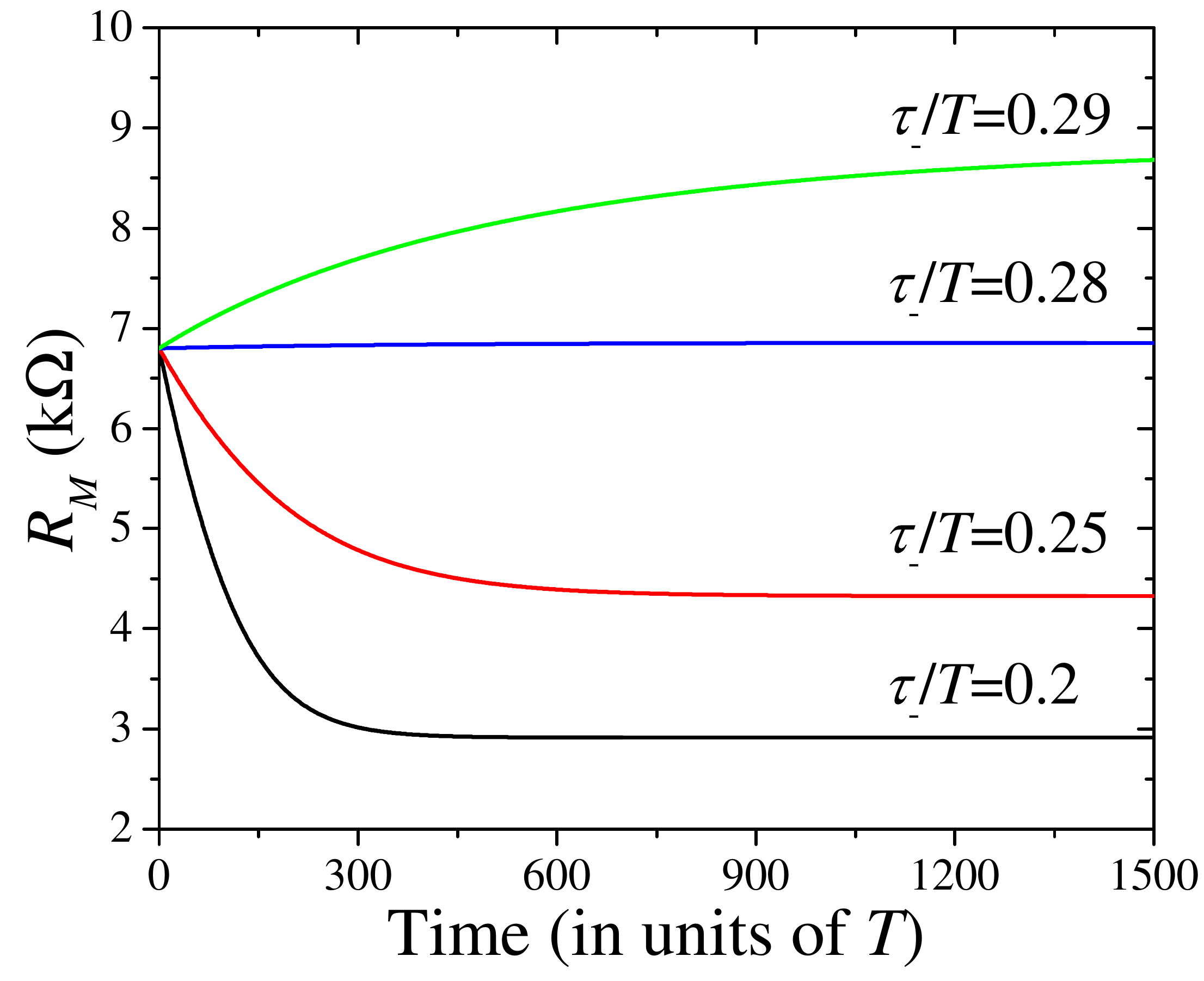}
\caption{ (a) Evolution of different initial states in the resistor-memristor circuit. The same final state for a variety of initial conditions indicates the attractor point in the memristor dynamics.
(b) Pulse-width control of the attractor point.
This figure was obtained using the same model parameters  as in Fig.~\ref{fig:2} and $kT=0.05$~V$^{-1}$.}
\label{fig:3}
\end{figure}


In the resistor-memristor network, the applied voltage is divided between the memristor and in-series connected resistor. Consequently,
the voltage across the memristor depends on its state and thus the state of memristor influences the memristor dynamics. Therefore, the entire network can be considered as an effective memristor with state-dependent dynamics. Below, we consider the network employing a threshold-type memristor (Eqs. (\ref{eq:30}) and (\ref{eq:33}) model) and find the attractor solution in a closed analytical form.

For the sake of simplicity, let us assume that $|V_M|\geq |V_{on,off}|$ when a pulse is applied.
In this case, for all possible states of memristor,  Eq. (\ref{eq:3}) can be presented as
\begin{equation}\label{eq:31}
  \frac{R_M(x_a)}{R+R_M(x_a)}\kappa-p=0,
\end{equation}
where
\begin{eqnarray} \label{eq:31a}
\kappa=V_{+}\tau_++V_{-}\tau_-,\\
p=V_{on}\tau_++V_{off}\tau_-.
\end{eqnarray}
Moreover, Eq. (\ref{eq:5}) leads to
\begin{equation}\label{eq:32}
\frac{R}{(R+R_M(x_a))^2}R'_M\kappa<0.
\end{equation}
As $\textnormal{d} R_M/\textnormal{d} x= R_{on}-R_{off}<0$, the necessary condition for the existence of attractor is
$\kappa=V_+\tau_++V_-\tau_->0$.

Next, we derive the potential function of the resistor-memristor network. Substituting Eq. (\ref{eq:30}) into Eq. (\ref{eq:6}), the following expression is obtained:
\begin{equation} \label{eq:35}
U(x)=\frac{kR\kappa}{R_{on}-R_{off}}\ln\left[R+R_M(x) \right]-k(\kappa-p)x.
\end{equation}
The potential function minima are defined by  $\textnormal{d} U(x)/\textnormal{d} x=0$ and $\textnormal{d}^2 U(x)/\textnormal{d} x^2>0$
resulting in the above Eqs. (\ref{eq:31}) and (\ref{eq:32}). The solution of Eq. (\ref{eq:31}) is
\begin{equation} \label{eq:36}
  R_M(x_a)=R\frac{p}{\kappa-p}.
\end{equation}
Additionally, we require that the attractor is located between $R_{on}$ and $R_{off}$, namely,
\begin{equation}\label{eq:37}
  R_{on}<R\frac{p}{\kappa-p}<R_{off}.
\end{equation}

The potential function of the resistor-memristor network is exhibited in Fig. \ref{fig:2} for several values of $\tau_-/T$. This plot demonstrates the
existence of a single attractor point in agreement with the above analysis. The position of attractor (the minimum of $U(x)$) shifts to the left with
$\tau_-$. According to Fig. \ref{fig:2}, the attractor is located  at $x\approx 0.9$, 0.7 and 0.4 for $\tau_-/T=0.2$, 0.25 and 0.28, respectively. Alternatively, the attractor position may be controlled by varying $V_{+}$ or $V_{-}$ instead of the pulse width~$\tau_-$.



Dynamics of resistor-memristor network was also simulated numerically. For this purpose, Eq. (\ref{eq:30}) was numerically integrated taking into account the voltage division between the resistor and memristor. Our numerical results are in perfect agreement with the analytical findings. First of all, the existence of attractor point is presented in Fig. \ref{fig:3}(a), which demonstrates that the distinct initial states of memristor converge to the same final state. Second, in Fig. \ref{fig:3}(b) it is shown that the attractor point can be controlled by the pulse width. We emphasize that the attractors in Fig. \ref{fig:3}(b) are located precisely at the minima of the memristor potential function in Fig. \ref{fig:2} (note that the same parameters were used in both calculations).

\subsection{Resistor-two memristors network} \label{sec:44}

\begin{figure}[tb]
\centering  \includegraphics[width=.9\columnwidth]{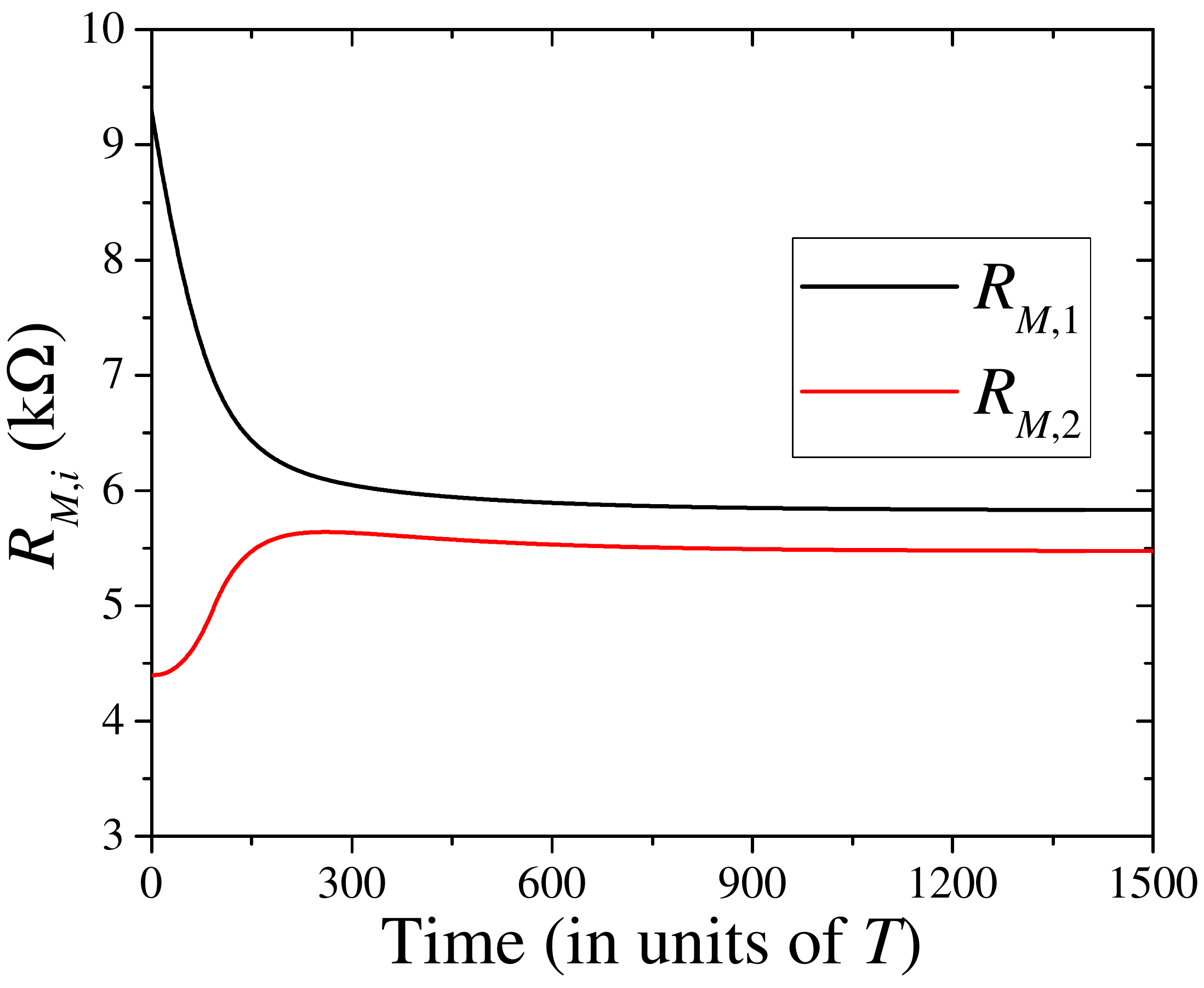}
\caption{Evolution of device states in the resistor-two memristors network. The calculation parameters are given in the text.}
\label{fig:4}
\end{figure}

Consider a network of in-series connected resistor and two threshold-type memristors. Using numerical simulations,
the network dynamics was investigated for several values of system parameters. Attractor solutions were observed for memristors connected with the same polarity. An example of network dynamics is shown in Fig. \ref{fig:4} found using $R=3$ $k\Omega$, $R_{on}=3$ $k\Omega$, $R_{off}=10$ $k\Omega$, $V_{on}=1$ V, $V_{off}=-0.7$ V (memristor 1),  $V_{off}=-0.8$ V (memristor 2), $V_{+}=-V_-=3$ V, $\tau_+/T=0.4$, $\tau_-/T=0.17$,  $kT=0.05$ V$^{-1}$ (memristor 1),  $kT=0.06$ V$^{-1}$ (memristor 2). The same parameter values were employed to obtain Fig. \ref{fig:1}(c) showing the network trajectories in the memristances space for selected initial conditions.

To find attractor points analytically, we use the theory developed in Sec. {\it Multivariable Memristors} for multivariable memristors.
It is not difficult to show that, indeed, the resistor-two memristors network can be described as a second-order memristive system
(Eqs. (\ref{eq:1}) and (\ref{eq:2}) with $n=2$).
 By solving Eq. (\ref{eq:8}) for our network (note that Eq. (\ref{eq:8}) is a vector analog of Eq. (\ref{eq:31})), we identify a possible attractor point
\begin{equation} \label{eq:41}
  R_{M,i}=R \frac{p_i}{\kappa-p_1-p_2},
\end{equation}
where
\begin{equation} \label{eq:42}
p_i=V_{on}^{(i)}\tau_++V_{off}^{(i)}\tau_-,
\end{equation}
and $i=1,2$ is the index of memristor.

There are two conditions that must be satisfied so that the solution given by Eq.(\ref{eq:41}) is an attractor:
(1) The obvious requirement of $R_{on}< R_{M,i}<R_{off}$, and (2) the positive-definiteness of the $2\times 2$ matrix $\tilde{F}_{ij}$ (see Eq. (\ref{eq:18})). These result in two inequalities:
\begin{eqnarray} \label{eq:43}
\kappa &>& 0,
\\ \label{eq:44}
4k_1k_2(\kappa-p_1)(\kappa-p_2) &>& (k_1p_1+k_2p_2)^2.
\end{eqnarray}
While the first inequality (\ref{eq:43}) coincides with the corresponding condition for the resistor-memristor network analyzed in the subsection {\it Resistor-memristor network} , the second inequality (\ref{eq:44}) puts some extra limitations on the pulse parameters and network in the case of two nonidentical memristors.

Our analytical theory is in excellent agreement with the results of numerical simulations. In particular, using the same parameter values as those used to obtain Fig. \ref{fig:4}, employing Eq.~(\ref{eq:41}), we find that the attractor is located at $x_1=0.5980$ and $x_2=0.6483$
corresponding to $R_{M,1}=5.814$ k$\Omega$ and $R_{M,2}=5.462$ k$\Omega$ perfectly matching the attractor point in Figs.  \ref{fig:1}(c) and  \ref{fig:4}. A direct check confirms that the inequalities  (\ref{eq:43}) and (\ref{eq:44}) are satisfied at the attractor point.

\begin{figure}[t]
\centering \includegraphics[width=60mm]{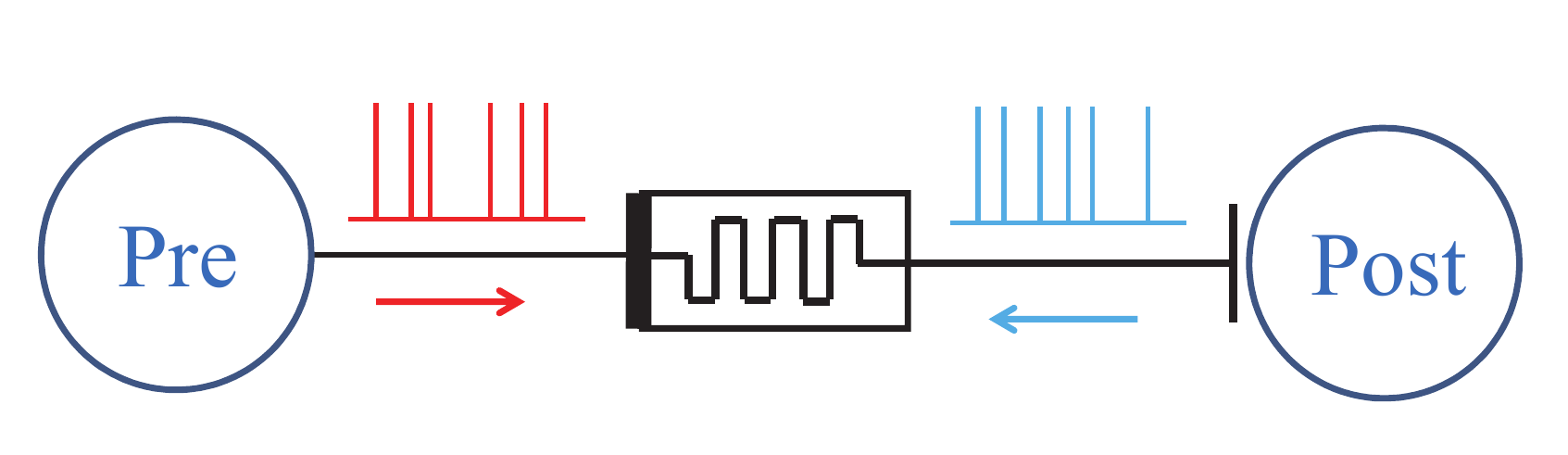}
\caption{Memristive synapse subjected to forward- and back-propagating spikes.}
\label{fig:5}
\end{figure}

\section{Conclusion} \label{sec:5}

In summary, the possibility of stable fixed points in the time-averaged dynamics of pulse-driven memristors and their networks has been demonstrated.
The necessary and sufficient conditions for fixed-point attractors have been derived. It has been suggested that some of driven memristors can be described by a potential function, which is an unexpected finding by itself. Several examples have been considered, and attractor states have been identified in the dynamics of second-order memristors and simple memristive networks. Our findings can be used to tune the resistance of analog memristors, and improve the models thereof.

There are strong indications in the literature that the attractor states play a significant role in biological neural dynamics ~\cite{cossart2003attractor,LEUTGEB2005345,Miller16a}. Our results indicate that a similar behavior can be realized in artificial neural networks with memristive synapses subjected to forward- and back-propagating spikes~\cite{pershin09c} (see Fig. \ref{fig:5}). This provides an additional argument in favor of using memristors in artificial neural networks. We finally note that our results are relevant to various experimental situations~\cite{gerasimova2017,ignatov16a,Ascoli16a}, and can be transferred to other promising systems with memory such as memcapacitors and meminductors~\cite{diventra09a}.



\bibliographystyle{eplbib}
\bibliography{memcapacitor}

\end{document}